\documentclass[a4paper,11pt]{article}
\pdfoutput=1 

\usepackage{jinstpub} 

\title{\boldmath Dynamical Chaos and Level Splitting under the Channeling of the High Energy Positrons in [100] Direction of the Silicon Crystal}

\author[a,1]{V.V. Syshchenko,\note{Corresponding author.}}
\author[a]{A.I. Tarnovsky,}
\author[a]{V.I. Dronik}
\author[b]{and A.Yu. Isupov}

\affiliation[a]{Belgorod State University,\\Pobedy Street, 85, Belgorod 308015, Russian Federation}
\affiliation[b]{Laboratory of High Energy Physics (LHEP), Joint Institute for Nuclear Research (JINR),\\Dubna 141980, Russian Federation}

\emailAdd{syshch@yandex.ru}

\abstract{The motion of charged particles in a crystal in the axial channeling regime can be both regular and chaotic. The chaos in quantum case manifests itself in the statistical properties of the energy levels set. These properties have been studied previously for the electrons channeling along [110] direction of the silicon crystal, in the case when the classical motion was completely chaotic, as well as for the ones channeling along [100] direction, when the classical motion can be both regular and chaotic for the same energy depending on the initial conditions. Here we study the positrons channeling in [100] direction. This case is of special interest due to the substantial tunneling probability between dynamically isolated regular motion domains in the phase space. The interaction of the energy levels via tunneling distinctly changes the level spacing statistics. All transverse motion energy levels as well as corresponding stationary wave functions are computed numerically for the 30 GeV positrons channeling in [100] direction of the silicon crystal. The values of the matrix elements for the tunnel transitions are extractad from these data. These results confirm the chaos assistance for the tunneling and the level splitting. These values will be used in the further researches of the quantum chaos manifestations in the channeling phenomenon.}


\arxivnumber{2310.10439} 

\proceeding{XIV International Symposium ``Radiation from Relativistic Electrons in Periodic Structures''\\
  September 18-22, 2023\\
 Tsaghkadzor, Armenia}

\begin{document}
\maketitle
\flushbottom

\section{Introduction}
\label{sec:intro}

When a fast charged particle is incident on a crystal at a small angle to any crystallographic axis densely packed with atoms, it can perform the finite motion in the transverse plane. This motion is known as the axial channeling \cite{AhSh, AhSh2, Ugg}. The particle motion in this case could be described with a good accuracy as the one in the continuous potential of the atomic string. During motion in this potential the longitudinal particle momentum $p_\parallel$ is conserved, so the motion description is reduced to two-dimensional problem of motion in the transversal plane. From the viewpoint of the dynamical systems theory, the channeling  particle's motion could be either regular or chaotic. The quantum chaos theory \cite{9, Reichl, Bolotin.book} predicts qualitative differences for these alternatives. 

The manifestations of chaos in quantum systems are found, first of all, in the statistical properties of their energy spectra. The quantum chaos theory predicts (see, e.g., \cite{9, Reichl, Bolotin.book}) that the distances between subsequent energy levels follows Wigner distribution formula if the quantum system obeys chaotic dynamics in the classical limit while the level spacings of the integrable (regular in the classical limit) system follows the exponential formula one (characteristic to Poisson flow). The case of the electrons channeling near [110] direction in silicon crystal has been studied in \cite{Pov.2015, NIMB.2016, Pov.2021}. In that case each pair of the closest parallel atomic strings forms the two-well potential (see, e.g., \cite{AhSh2}), in which the motion above the saddle point is chaotic for the major part of the initial conditions. As a result, the level spacing statistics in that case is well described by Wigner distribution.

More complex and interesting is the case of co-existence of the domains of regular and chaotic dynamics in the system's phase space, such situation takes the place when the electron or positron channels near $[100]$ direction of silicon crystal. The first approach to description of this situation was made by Berry and Robnik \cite{Berry}. These authors have assumed that the spectrum of such mixed system consists of two independent level sequencies: one is generated by all regular motion domains, and another is generated by (the only) chaotic domain. This leads to the level spacing distribution with the only parameter --- the relative contribution $\rho_1$ of the regular domains to the semiclassical mean level density. The estimation procedure for this value is described in \cite{JINST.2019}, where the Berry--Robnik distribution \cite{Berry} was found to be satisfactory for the level spacing statistics of the electrons channeling in $[100]$ direction.

However, the levels in the regular and chaotic sequences are not independent: the interaction between levels changes the level spacing statistics in a way demonstrated by Podolskyi and Narimanov in the series of papers \cite{Narimanov.1, Narimanov.2, Narimanov.3}. They stated that the presence of the chaotic motion domain can increase the tunneling rate between the dynamically isolated regular domains in the phase space (the so-called chaos-assisted tunneling, CAT \cite{Narimanov.1}) and obtained the formula that accounts this influence on the level spacing statistics. The Podolskyi--Narimanov distribution \cite{Narimanov.3} needs two parameters: the value $\rho_1$ same as for Berry--Robnik distribution, and the mean amplitude for the tunneling transitions. The aim of the present paper is to find the transition amplitudes for all interacting energy levels for the transverse motion of the 30 GeV positrons channeling in the $[100]$ direction of the silicon crystal. The results agree with the expectations for the CAT mechanism and could be used for the testing of the  Podolskyi--Narimanov prediction.

\section{Method and potential well}
\label{sec:method}

\begin{figure}[htbp]
\vspace*{-2mm}\centering
\includegraphics[width=0.45\textwidth]{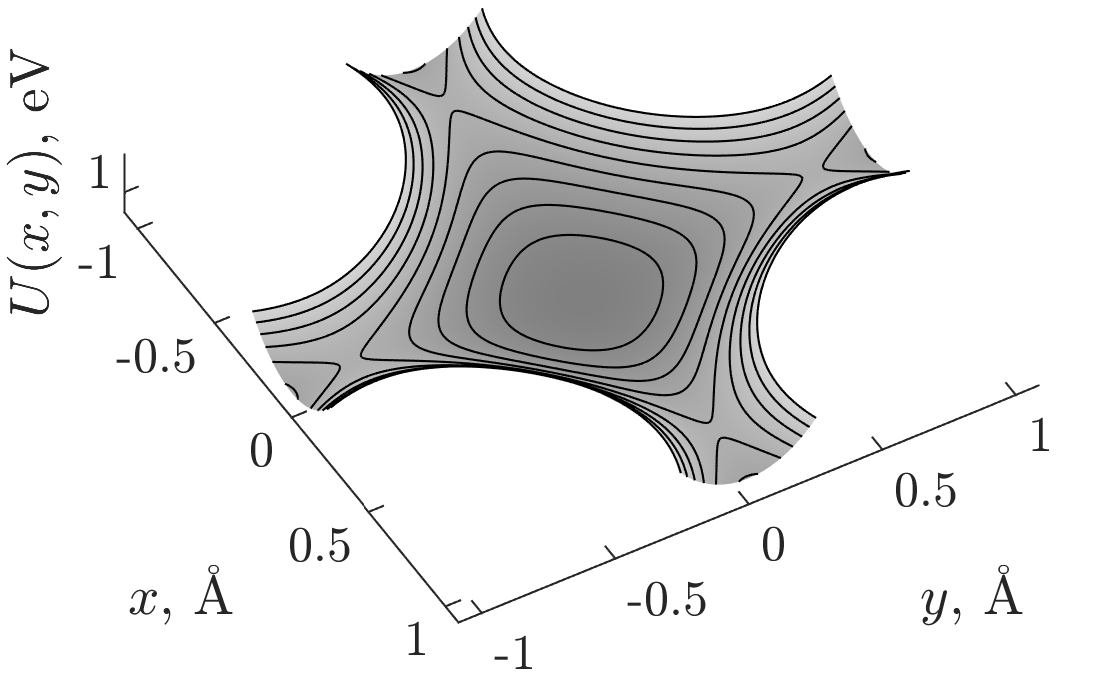} 
\vspace*{-2mm}\caption{\label{potentials} Potential (\ref{U.pos}).}
\end{figure}

The particle's transversal motion in the atomic string continuous potential is described by the two-dimensional Schr\"odinger equation with Hamiltonian
\begin{equation}
    \label{Hamiltonian}
    \hat{H} = - (c^2\hbar^2 / 2 E_\parallel)  \left[ (\partial^2/\partial x^2) + (\partial^2/\partial y^2) \right] + U(x, y) \,,
\end{equation}
where the value $E_{\parallel} / c^2$ (here $E_{\parallel} = (m^2
c^4 + p_{\parallel}^2 c^2)^{1/2}$) plays the role of the particle mass
\cite{AhSh}. The Hamiltonian eigenvalues $E_\perp$ as well as the eigenfunctions corresponding to them are found numerically using the so-called spectral method \cite{3, Dabagov3, NIMB.2013}. Here we consider the positron motion near direction of the atomic string $[100]$ of the Si crystal. The continuous potential could be represented by the modified Lindhard potential
\cite{AhSh}
\begin{equation}
    \label{U.1}
    U^{(1)} (x, y) = U_0 \ln \left[ 1 + \beta R^2 / (x^2 + y^2 + \alpha R^2) \right] ,
\end{equation}
where $U_0 = 66.6$~eV, $\alpha = 0.48$, $\beta = 1.5$, $R =
0.194$~\AA~(Thomas--Fermi radius). These atomic strings form in the plane $(100)$ the square lattice with the period $a \approx 1.92$ \AA .  Axial channeling of positrons near $[100]$ direction is possible in the  small potential pit near the center of the square cell with repulsive potentials $U^{(1)}$ in the corners of the square (Figure \ref{potentials}):
\begin{equation}
    \label{U.pos}
    U (x, y) = U^{(1)} (x-a/2, y-a/2 )
    +U^{(1)} (x-a/2, y+a/2) + \vphantom{a}
\end{equation}
\begin{equation*}
   U^{(1)} (x+a/2, y-a/2)
    +U^{(1)} (x+a/2, y+a/2) - 7.9589\ \mathrm{eV} ,
\end{equation*}
where the constant is chosen to achieve zero potential in the center of the cell. The spectral method for the channeling electrons and positrons near $[100]$ direction had been tested in \cite{group} for small $E_\parallel$ values, when the total number of energy levels in the potential well is small. Here we put $E_\parallel = 30$~GeV to achieve the semiclassical domain, where the energy level density is high, as it is needed for quantum chaos investigations. The obtained level density is about 150 levels (for every of 4 types of the eigenstates symmetry, see below) per 1 eV in the upper part of the potential well (in the interval marked by the yellow band in Figure \ref{density}). 

The channeling particle's eigenstates of the transverse motion can be classified using the group theory. The
potential (\ref{U.pos}) possesses the symmetry of the square thus isomorphic to the group $C_{4v}$ (or  $D_4$). This group has four one-dimensional irreducible representations and one two-dimensional one, denoted
$A_1$, $A_2$, $B_1$, $B_2$ and $E$ \cite{group, LL3}. So, the eigenstates of the first four types are non-degenerated while the eigenstates of the last type are twice degenerated and are not of our interest. 

\begin{figure}[htbp]
\vspace*{-2mm}\centering
\includegraphics[width=\textwidth]{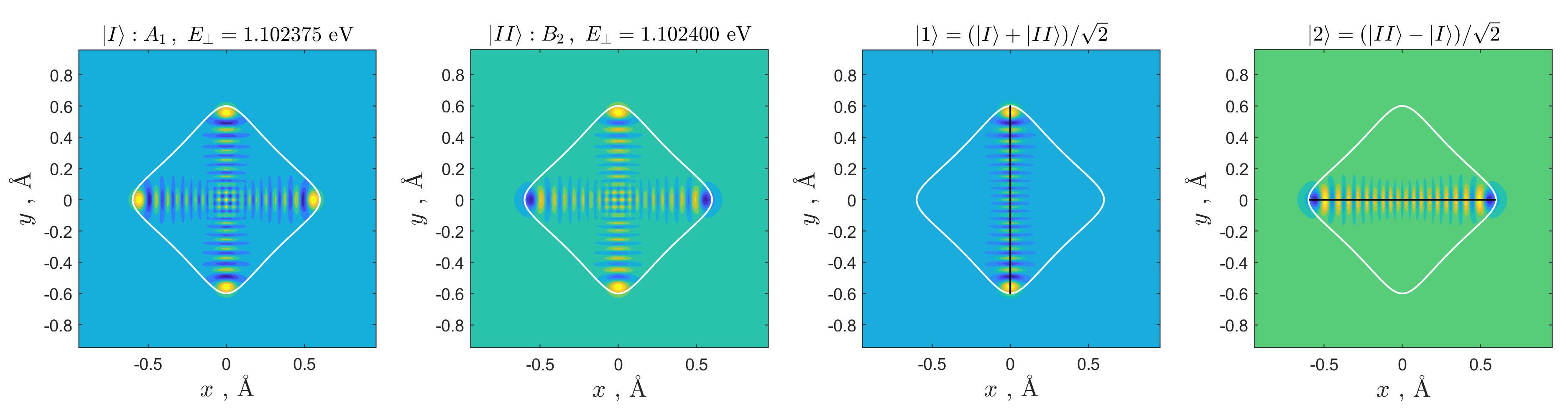}  \\
\includegraphics[width=\textwidth]{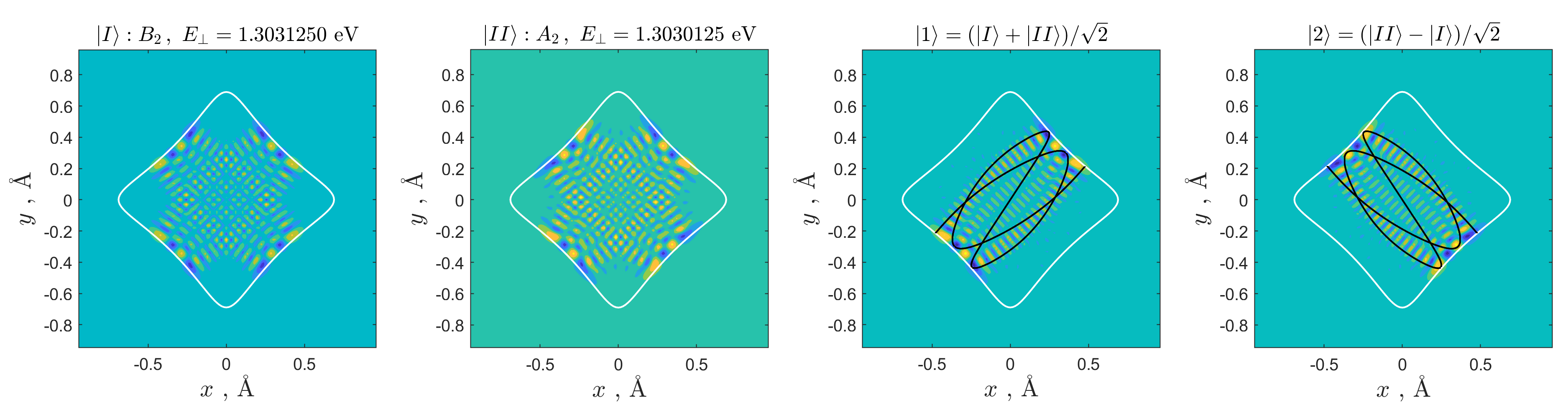}
\vspace*{-10mm}\caption{\label{pare.of.states.fig} Two examples of pairs of stationary states with close energies (\ref{pare.of.states}) (\emph{left}) and their linear combinations with clear semiclassical meaning (\emph{right}).}
\end{figure}

\section{Results and discussion}

The mechanism of the levels interaction could be understood from the well-known theory of the two-level system described in many textbooks, see, e.g., \cite{Feyn.3} (volume 3, chapters 8--10). This theory starts from the Hamiltonian that neglects the possibility of the tunneling transition between two degenerated states $\left|1\right>$ and $\left|2\right>$ (corresponding to the energy $E$) with some semiclassical meaning. The inclusion of the tunneling possibility (which is described by nonzero amplitude of transition between these two states $\left<2\right| \hat{H} \left| 1\right> = V$) at the next step of the consideration leads to the interaction between these states and hence the mutual displacement of the levels that removes the degeneracy and forms the final resulting spectrum of the system's stationary states. So the stationary states with the definite energies $E+V$ and $E-V$ are the linear combinations of the states $\left|1\right>$ and $\left|2\right>$:
\begin{equation}\label{pare.of.states}
    \left|I\right> = \frac{\left|1\right> - \left|2\right>}{\sqrt{2}}\,, \quad 
    \left|II\right> = \frac{\left|1\right> + \left|2\right>}{\sqrt{2}} \,.
\end{equation}
This way is used in the case when we can solve the Schr\"odinger equation with the simplified Hamiltonian and after that can find the tunneling transitions amplitudes. The examples of such approach could be found in quantum mechanics textbooks (see, e.g., problem 3 in \S 50 \cite{LL3} or problem 9.17 in \cite{Griffiths}). The same situation is considered in \cite{Narimanov.1, Narimanov.2, Narimanov.3}: the authors start from the states which wave functions are localized in the dynamically isolated parts of the phase space and then take into account the tunneling possibility between them. Let us underline that only after that tunneling accounting the problem solution leads to the final set of the system's stationary states. In other words, the CAT formalism starts from the eigenstates that exist while neglecting the tunneling possibility. The tunneling changes the energy eigenvalues (that is interpreted as the level interaction) as wellas the level-spacing statistics. The final exact eigenstates do not tunnel anywhere.

\begin{figure}[htbp]
\vspace*{-3mm}\centering
\includegraphics[width=0.49\textwidth]{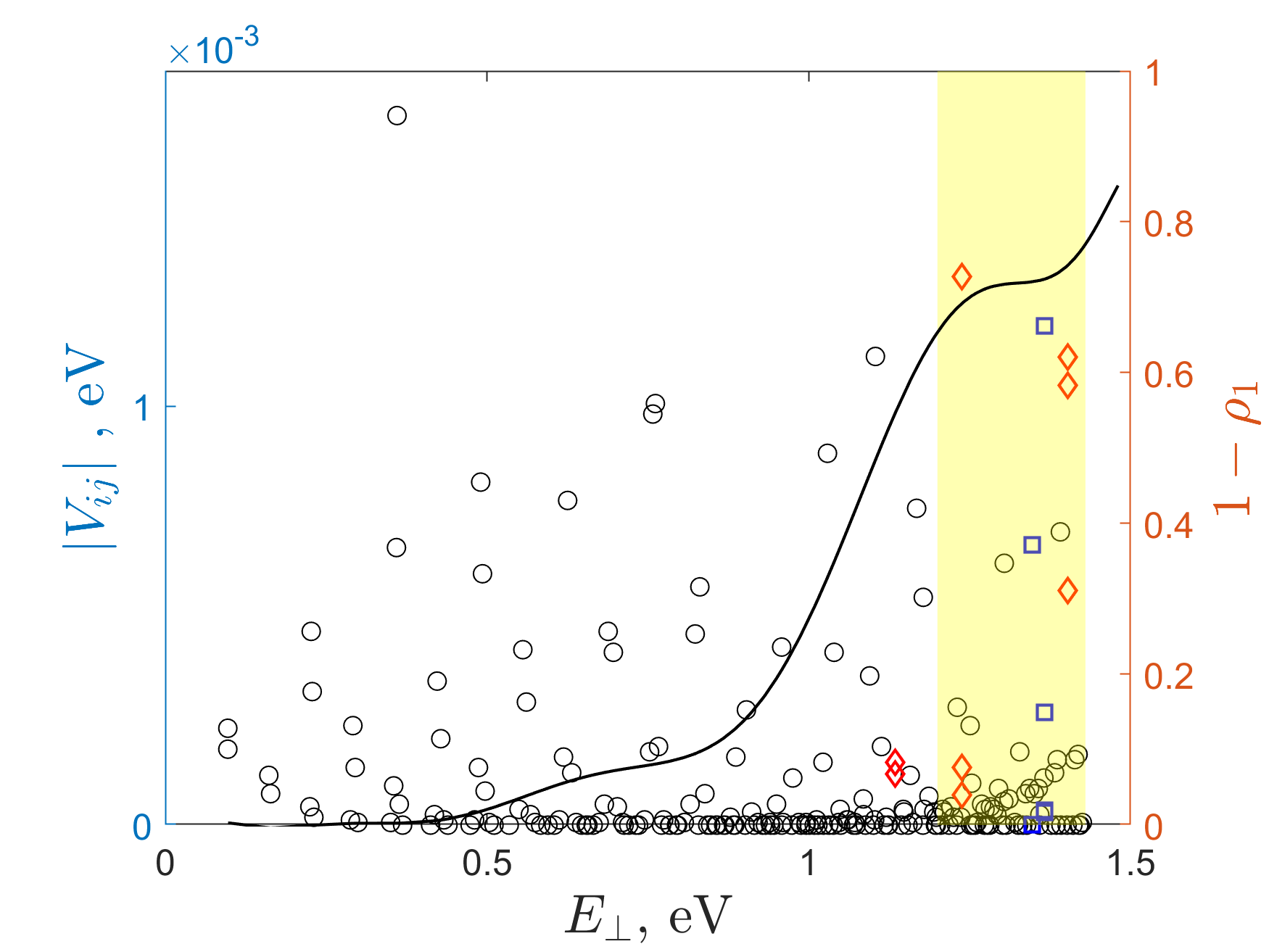} \ \ 
\includegraphics[width=0.49\textwidth]{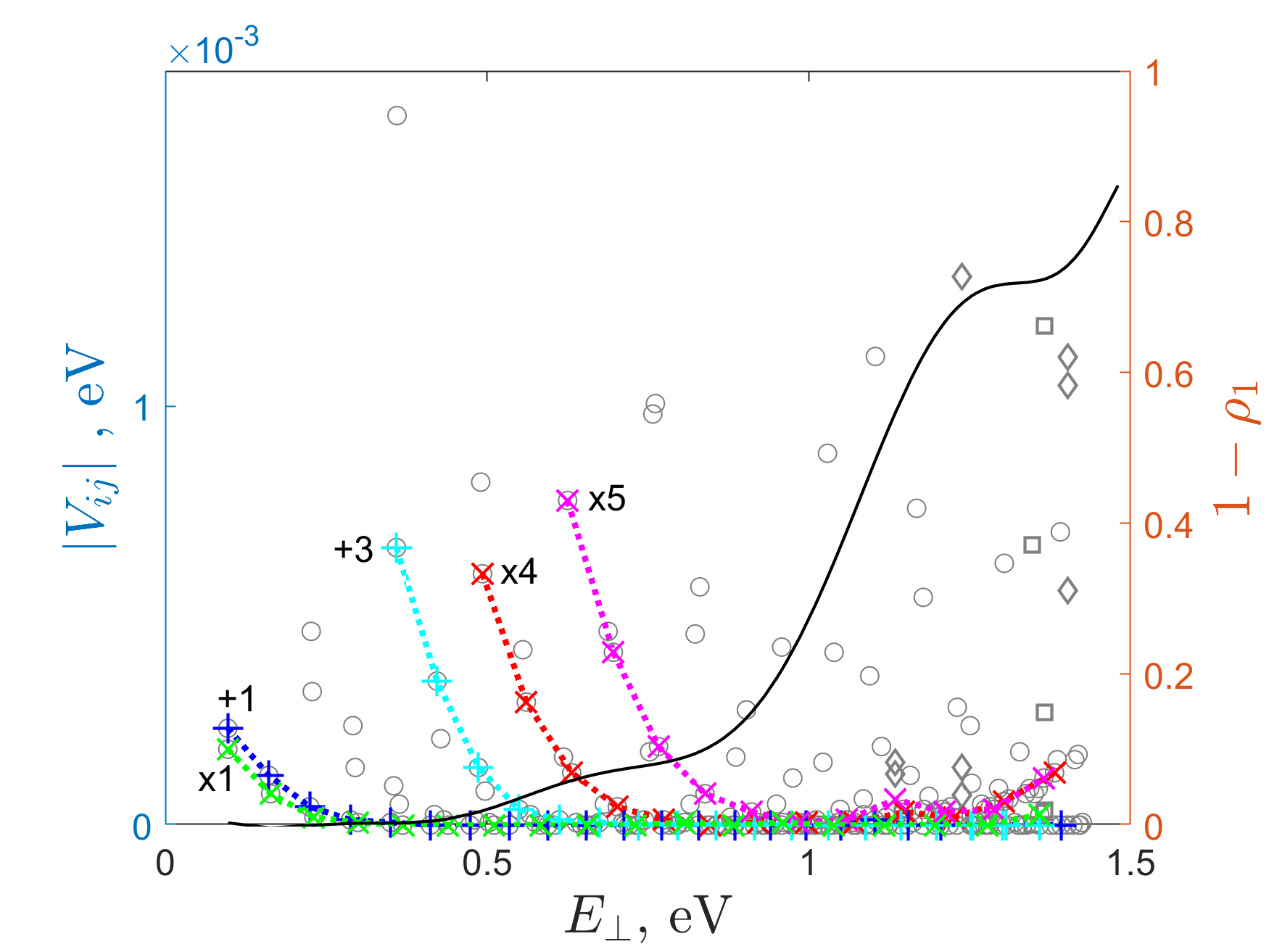} 
\vspace*{-9mm}\caption{\label{density} The absolute values of all transition amplitudes (circles, diamonds and squares, left scale) and relative contribution of the chaotic dynamics domain in the phase space to the semiclassical mean density of states (solid line, right scale).}
\end{figure}

In our paper we have the inverse problem: the numerical solution of the Schr\"odinger equation with the Hamiltonian (\ref{Hamiltonian}) gives us the exact set of the energy levels and the corresponding wave functions of the stationary states. And our goal is to reconstruct the tunneling transition amplitudes from these data. In fact, in \cite{Feyn.3} (volume 3, chapters 8--10) the same procedure has fulfilled: the authors reconstruct the phenomenological Hamiltonian 
parameters (the energy eigenvalue of two degenerated states and the tunneling amplitude between them) from the observed splitting of the levels. Two examples of pairs of stationary states (with close energies) along with their linear combinations are presented in Figure \ref{pare.of.states.fig}. The semiclassical meaning of these combinations is illustrated by the classical orbits pictured over the wave functions plots; we see that these classical trajectories are localized in the dynamically isolated regions and could not leak out classically from one to another. The absolute values of the transition amplitudes for all such pair of states found from the level splittings, $|V_{ij}| = |E_{\perp i} - E_{\perp j}|/2$\,,  are marked in Figure \ref{density} by circles.

\begin{figure}[htbp]
\vspace*{-5mm}\centering
\includegraphics[width=0.74\textwidth]{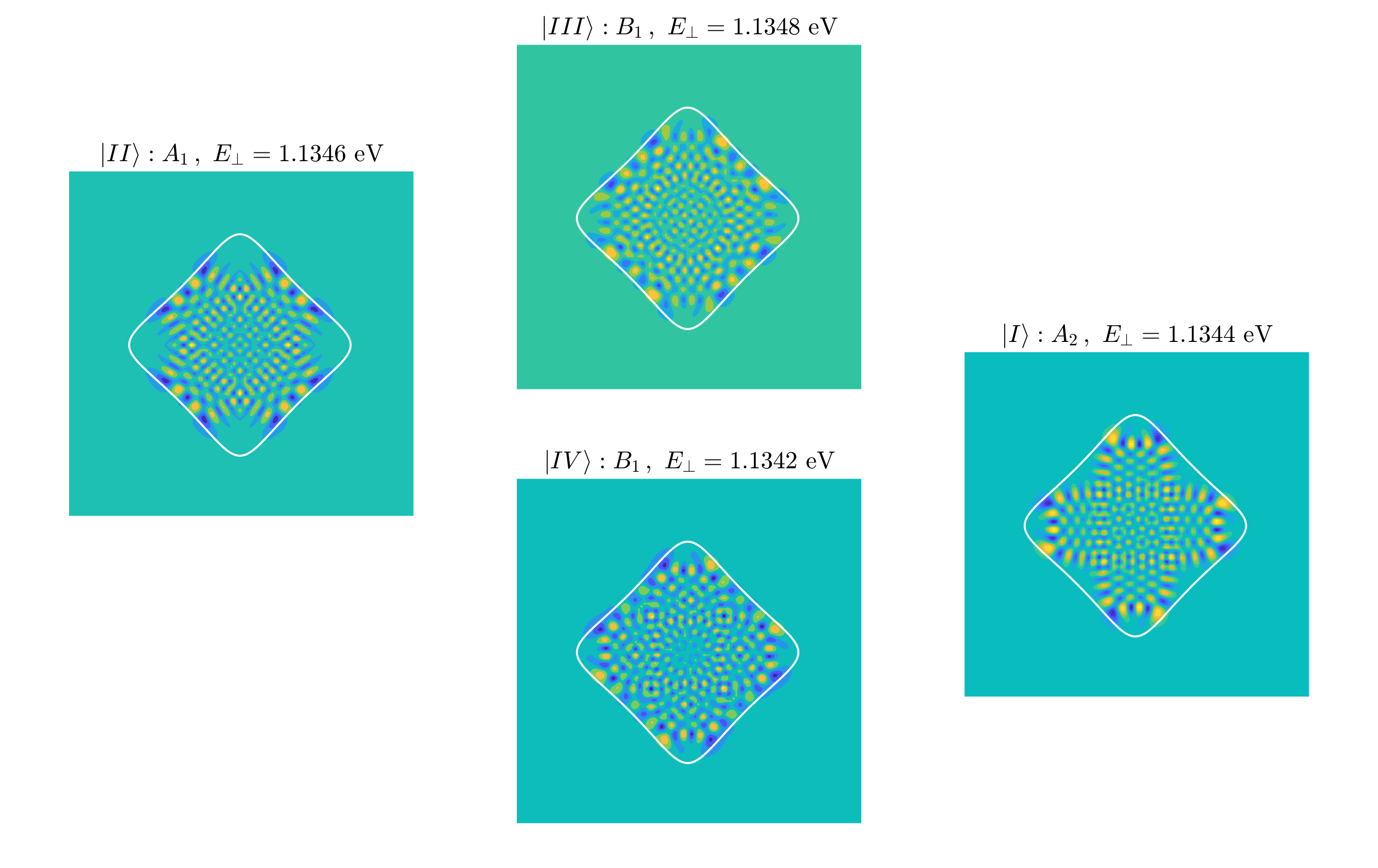} \ 
\includegraphics[width=0.24\textwidth]{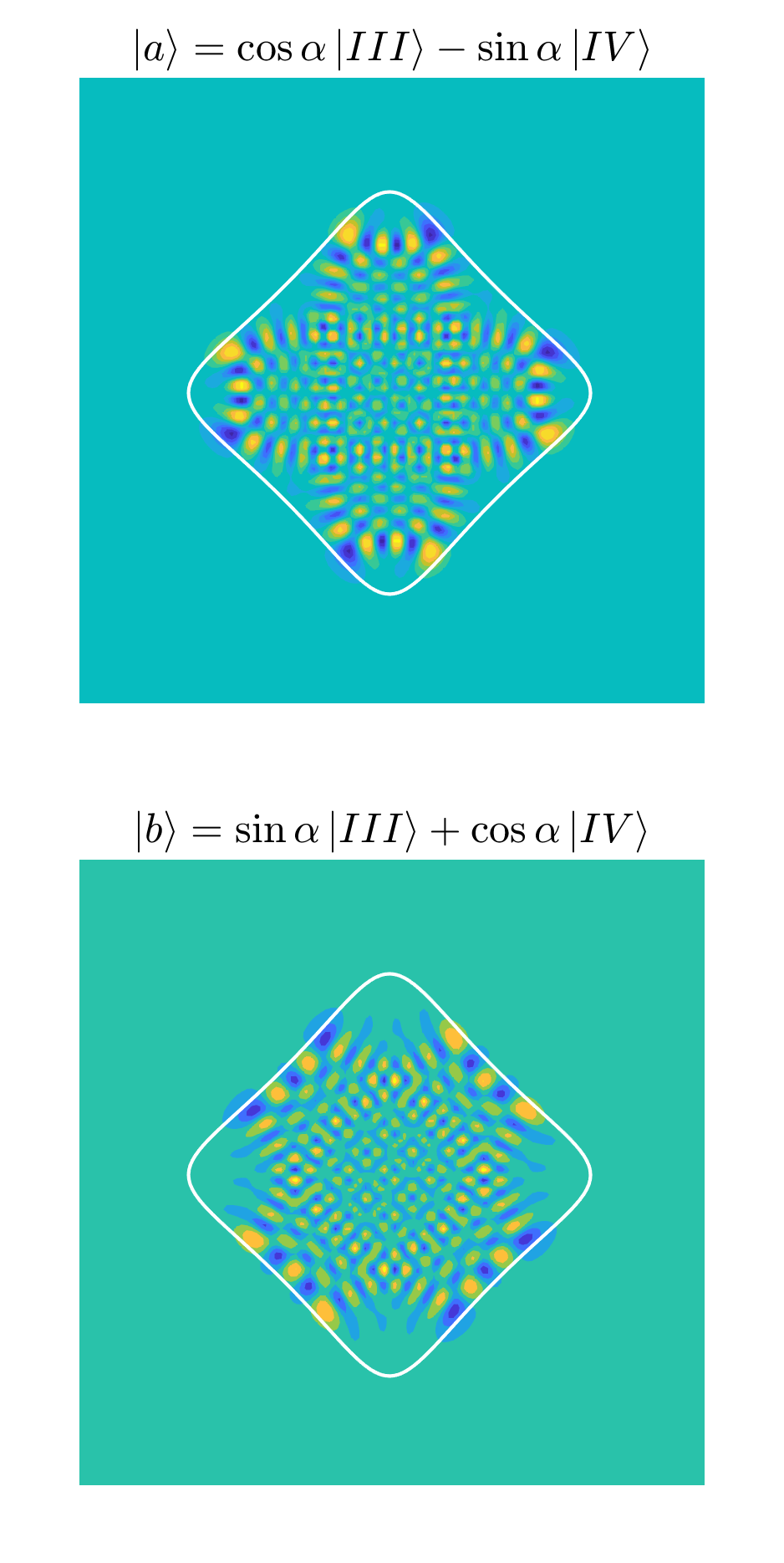} 
\vspace*{-7mm}\caption{\label{4.states} \emph{Left:} four eigenstates with close energies that belong to three different types of symmetry. \emph{Right:} the linear combination of the two eigenfunctions of $B_1$ type makes the functions that are counterparts to the another two eigenfunctions of the original quartet.}
\end{figure}

We found also some cases of the interaction between four energy levels. The linear combinations of the states $\left|III\right>$ and $\left|IV\right>$ from the example presented in Figure \ref{4.states} makes the couples to the states $\left|I\right>$ and $\left|II\right>$. The absolute values of such four-level interactions are marked in Figure \ref{density} by diamonds and squares.

The rule in the transition amplitude values could be seen if we classify the states according to the number of half-waves lying along the narrow side of the wave function. We distinguish the states that correspond to the classical orbits connecting two opposite corners (let us denote such states as $+$) from the states that correspond the orbits connecting two opposite sides (denote them as $\times$) of the potential well. So the states in Figure \ref{pare.of.states.fig} belong to the $+1$ and $\times4$ families in our classification while the states in Figure \ref{4.states} belong to $+6$ and $\times5$ families.

\begin{figure}[htbp]
\vspace*{-3mm}\centering
\includegraphics[width=0.49\textwidth]{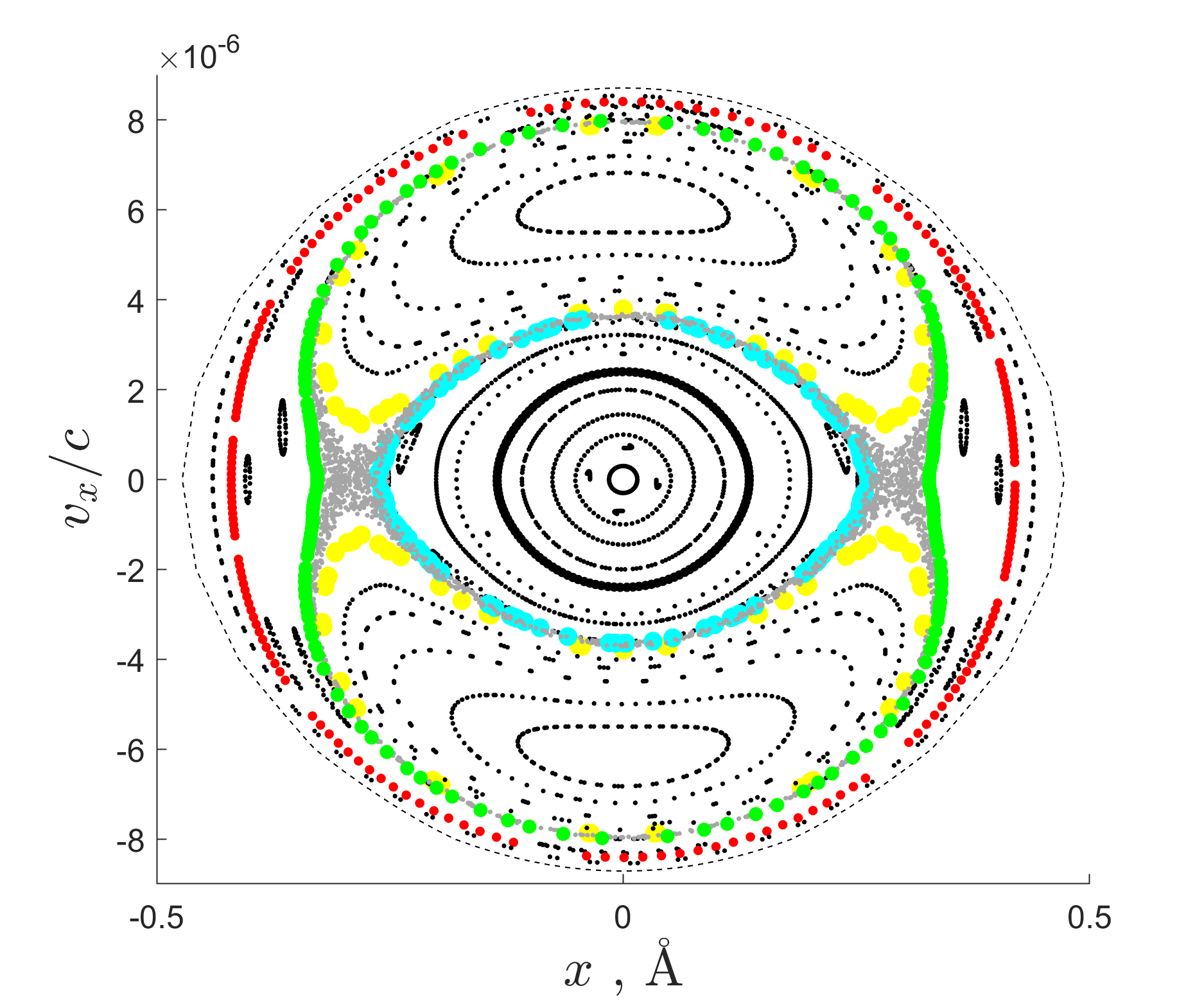} 
\includegraphics[width=0.49\textwidth]{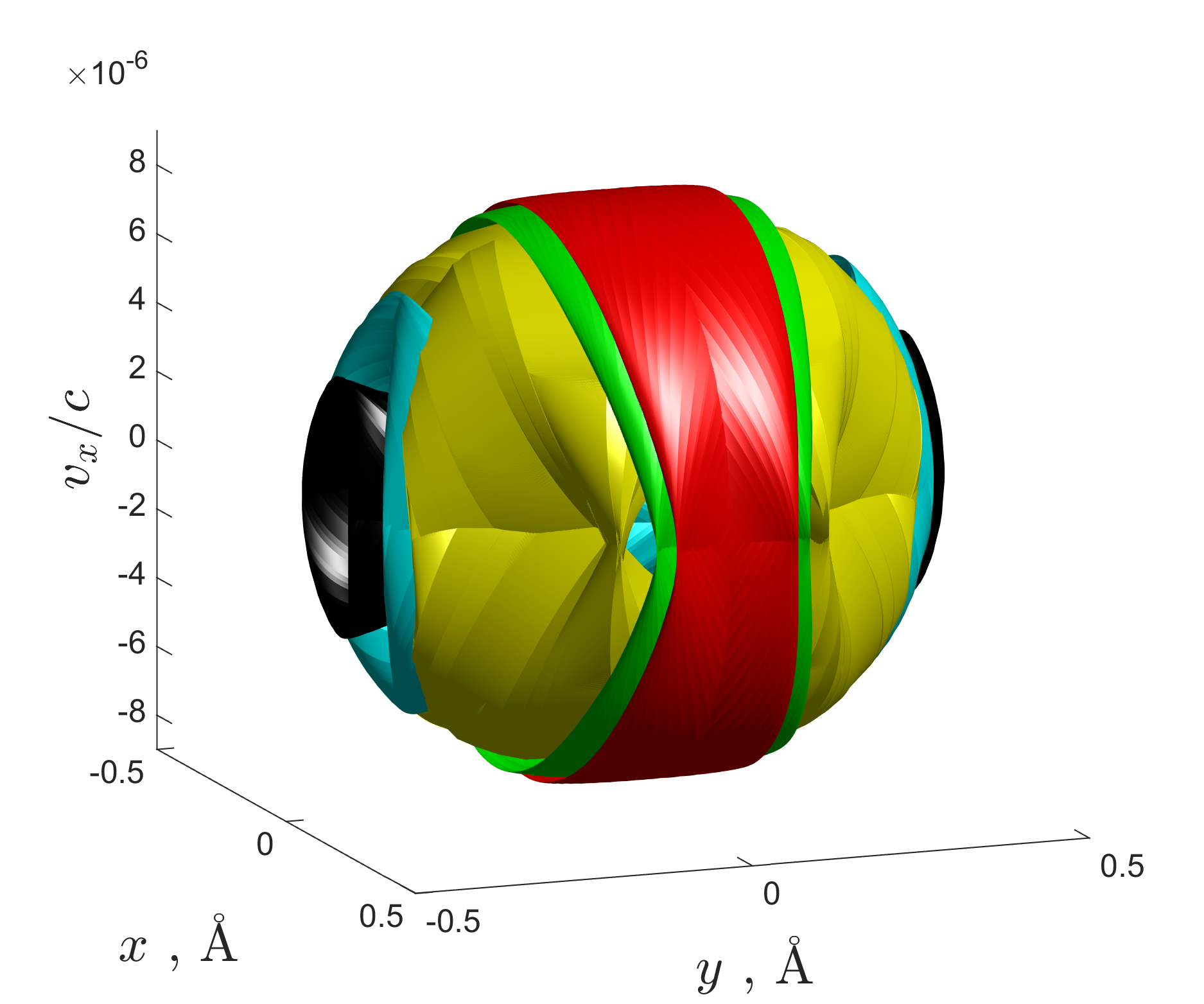} \\
\includegraphics[width=0.49\textwidth]{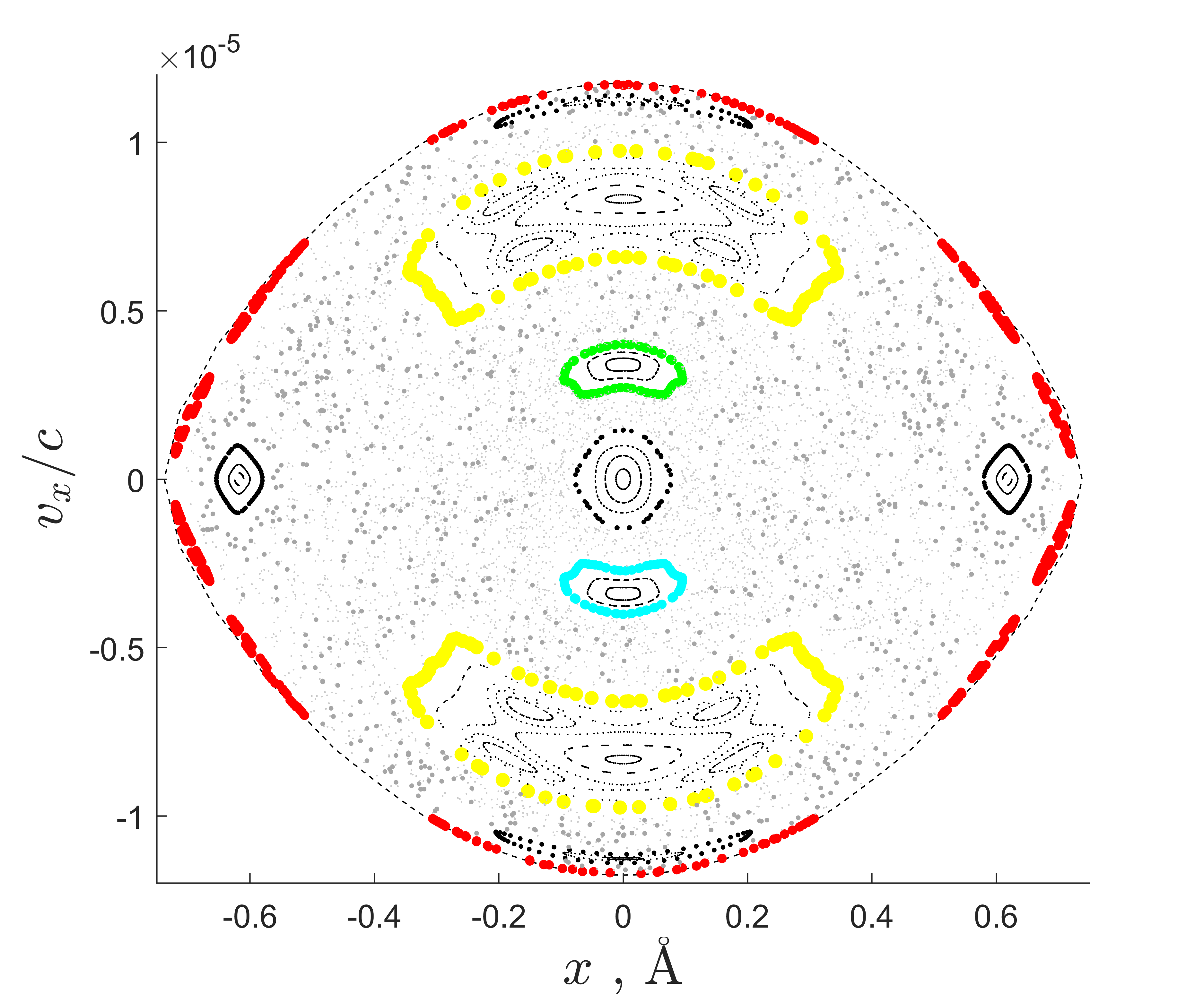} 
\includegraphics[width=0.49\textwidth]{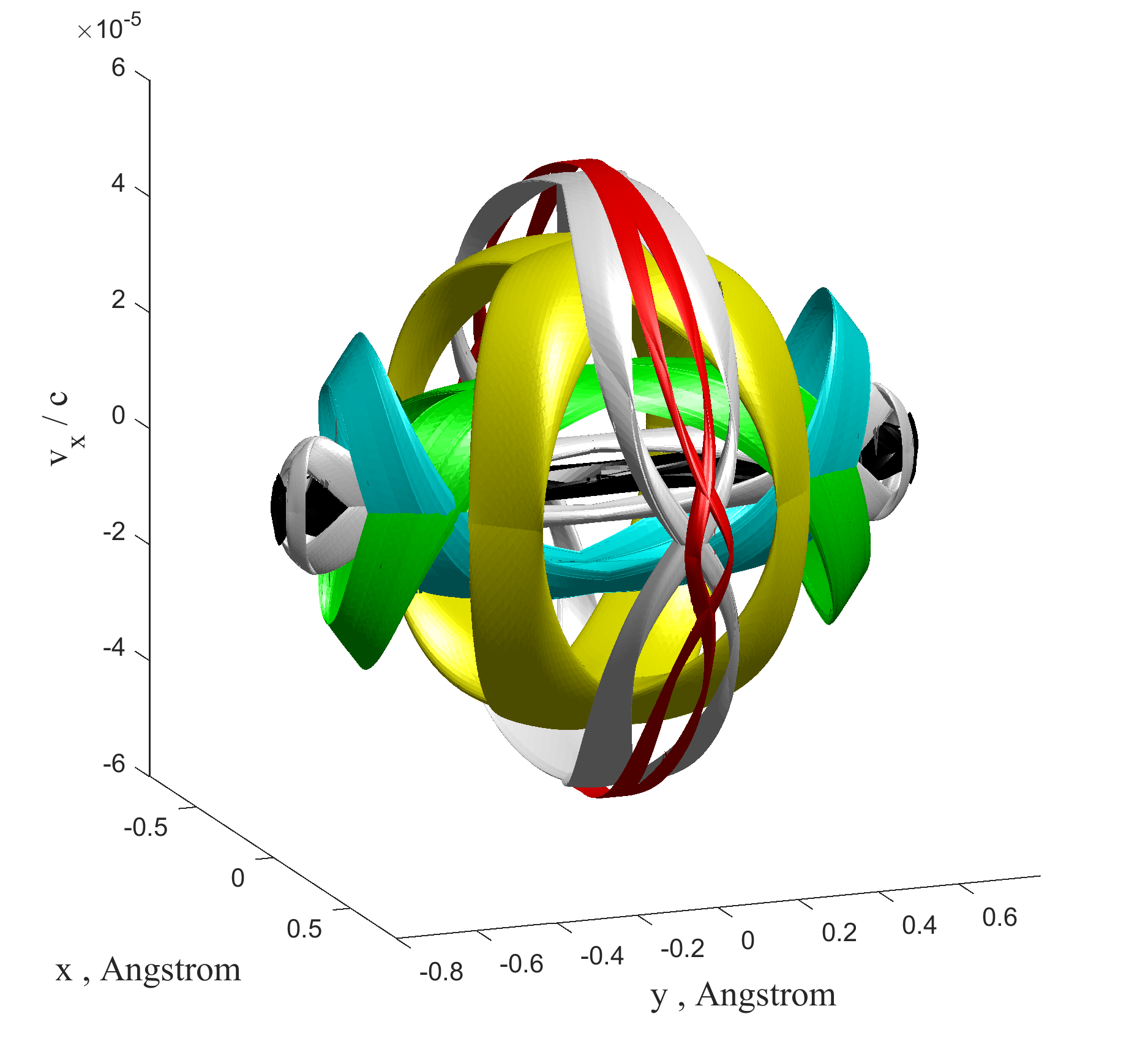} 
\vspace*{-3mm}\caption{\label{Poincare} Poincar\'e sections ({\it left}) and projections of the 4-dimensional regular dynamics domains to the subspace $(x, y, v_x)$ ({\it right}) for the transverse motion energy values 0.7687 eV {\it (upper row}) and 1.3865 eV ({\it lower row}).}
\end{figure}

Some of such families are merged by the color lines in the right panel of Figure \ref{density}, and we see some rule: the tunneling amplitude decreases while ascending from the bottom of the well and than increases again. The reason is in the relative contribution of the chaotic dynamics domain to the semiclassical mean density of states $1-\rho_1$ presented by the smooth line (related to the right scale) in this figures. The part of chaotic motion rapidly increases while the transverse motion energy approaches the upper edge of the well (compare upper and lower rows in Figure \ref{Poincare}).

\begin{figure}[htbp]
\vspace*{-2mm}\centering
\includegraphics[width=0.45\textwidth]{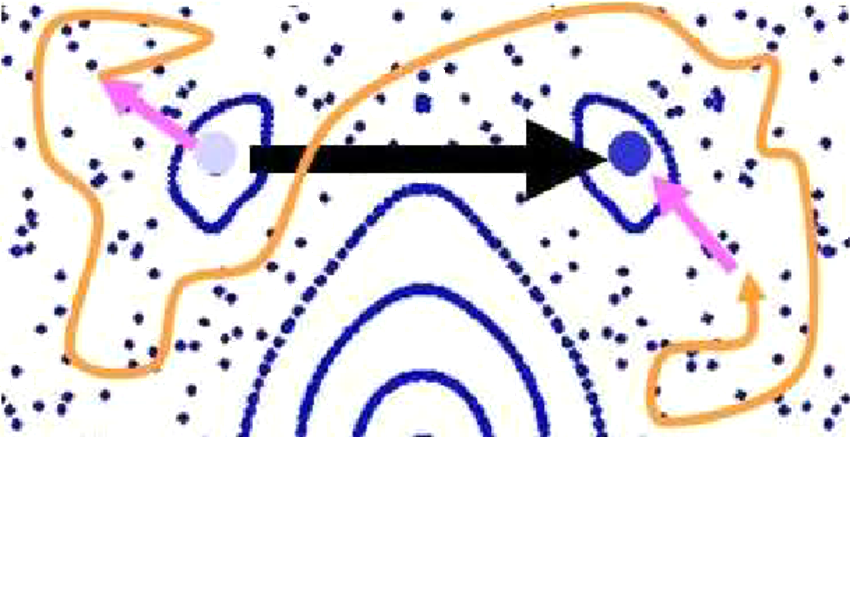} \ \ 
\includegraphics[width=0.48\textwidth]{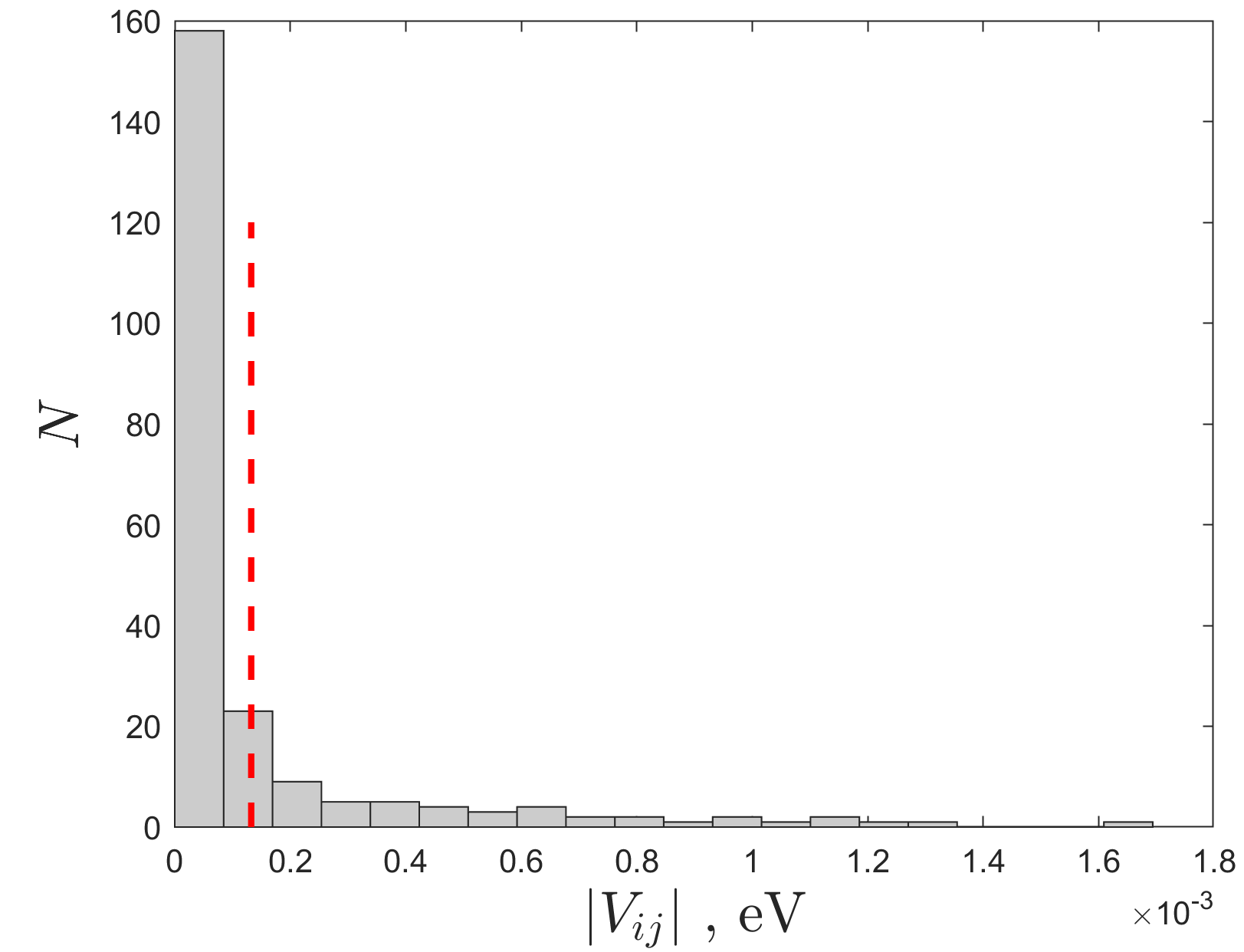}
\vspace*{-3mm}\caption{\label{CAT.and.result} \emph{Left:} Schematic \cite{Narimanov.1} indicating a direct tunneling process (black arrow) and a chaos-assisted tunneling process (yellow arrow) which would contribute to splitting of the degenerated levels. {\it Right:} Distribution of the found values of the tunneling amplitudes for the complete set of levels for 30 GeV channeling positrons. The dashed line correspond to the average value.}
\end{figure}

For every family of states the tunneling amplitude decreases and than increases again depending on the transverse motion energy. The particle dynamics is regular for almost all initial conditions in depth of the potential well, so the boundaries of the dynamically isolated regular domains touch each other. Hence, the direct tunneling would be highly probable if the wave function reaches the domain boundaries, that leads to the significant level splitting between symmetric and antisymmetric combinations (\ref{pare.of.states}). As the energy increases, the de Broglie wavelength decreases, therefore the wave functions with small number of halfwaves in ``short'' direction are found localized deeply inside the regular motion domain, surrounded by the regions of other regular-state localization. Therefore, the probability of tunneling through dynamically inaccessible regions decreases. However, as the energy continues to increase, the proportion of the phase volume corresponding to regular dynamics decreases. So, near the upper edge of the potential well, the phase space contains several regions dynamically isolated from each other and corresponding to different types of regular orbits separated by the region of chaotic dynamics. Therefore, wave functions of the selected family can again differ significantly from zero near the boundaries of the regions of regular motion. And although the width of the region that is dynamically inaccessible to particle motion is large, the probability of tunneling through it begins to increase due to the mechanism known as chaos-assisted tunneling (CAT). In this case the particle only needs to penetrate beyond its dynamically allowed region (with the amplitude $V_{RC}$), and it will be picked up by a chaotic flow (\cite{Narimanov.1}, Figure \ref{CAT.and.result}, left panel), which takes it to the border of the partner region sooner or later, where the particle will be able to successfully complete the tunneling process (see also the review of CAT studies in \cite{Bolotin.book}). This is the reason for the increase in level splitting near the well top.

\section{Conclusion}

The energy levels of the transverse motion of $E_\parallel = 30$ GeV and the corresponding eigenfunctions for the positrons channeling in the [100] direction of a silicon crystal are found using the spectral method of numerical integration of the time-dependent Schr\"odinger equation. Using these data we found the complete set of the tunneling transitions matrix elements for the tunneling between dynamically isolated semiclassical states. The dependence of their values on the transverse energy can be interpreted as a manifestation of the chaos-assisted tunneling. Also the obtained values of transition matrix elements could be used as the parameter in the Podolskyi--Narimanov distribution for the level spacings. That opens the new possibility for the testing of the predictions of the quantum chaos theory.

\end{document}